# Luminescence-Induced Tunable Superconductivity in BSCCO via GaP Quantum Dots

Qingyu Hai[*], Duo Chen[*], Ruiyuan Bi, Yao Qi, Lifeng Xun, Xiaoyan Li, Xiaopeng Zhao [**]

Smart Materials Laboratory, Department of Applied Physics, Northwestern Polytechnical University, Xi'an 710129, China;

haiqingyu@mail.nwpu.edu.cn(Q.H.); chenduo@mail.nwpu.edu.cn(D.C.); biruiyuan@mail.nwpu.edu.cn(R.B.); qiyao@mail.nwpu.edu.cn (Y.Q.); xunlifeng@mail.nwpu.edu.cn (L.X.);lixiaoyan0521@mail.nwpu.edu.cn (X.L.);

* These authors contributed equally to this work.

**Correspondence: xpzhao@nwpu.edu.cn

**Abstract:** The enhancement of superconducting properties in high-temperature copper-oxide superconductor B(P)SCCO remains a hot research topic in the field of superconducting materials. Building on previous research, here we introduce GaP quantum dots as an heterophase into the B(P)SCCO superconductor, aiming to enhance its superconductivity through the luminescent properties of GaP quantum dots. The experimental results demonstrate that the introduction of GaP quantum dots into B(P)SCCO generates significant tunable superconducting effects, leading to enhanced critical transition temperature ($T_c$), critical current density ($J_c$), and Meissner field ($H_c$) of B(P)SCCO with increasing luminescent intensity of the GaP quantum dots. The enhancement effect induced by GaP quantum dots exhibits a positive correlation with luminescent intensity, meaning samples with the addition of GaP quantum dots exhibiting higher luminescent intensity show elevated $T_c$, $J_c$, and $H_c$ values. Unlike impurity effects, a distinct critical concentration dependency is observed. Notably, this GaP quantum dot modification strategy is not only effective in conventional superconductors but also applicable to high-temperature oxide superconductors.

energy; smart superconductivity.

**1. Introduction**

Bismuth-based superconducting materials (general formula: $Bi_2Sr_2Ca_{n-1}Cu_nO_{2n+4+\delta}$) [1-5], as a representative copper oxide superconductor, exhibit significant application value in power transmission[6], high-field magnets[7], and energy storage due to their high critical temperature (Tc = 110K for Bi-2223 phase[5]), absence of rare-earth elements, and superior processability. However, the intrinsic limitations of BSCCO such as strong anisotropy caused by their layered structure[1, 8], weak interlayer coupling, and insufficient intrinsic flux pinning severely constrain performance in high-current and high-field environments[9-11]. Traditional elemental doping (e.g., Pb[12, 13], Sb[14, 15]) can improve phase purity, yet offers limited enhancement in Critical transition temperature (*Tc*) and critical current density (*Jc*). In contrast, the nanocomposite strategy — introducing heterophase nanoparticles as artificial pinning centers (e.g., insulator Al2O3[16], semiconductor SiC[17], ferromagnet FePb[18])—has emerged as an effective modification approach for (Bi,Pb)-Sr-Ca-Cu-O (B(P)SCCO) superconductors[19-28]. This strategy leverages interface coupling effects between nanoscale heterophases and the superconducting matrix to create strong pinning sites, significantly enhancing flux pinning force density (Fp) and boosting Jc by 1~2 orders of magnitude[29-33]. However, it typically induces Tc degradation. Clearly, although this method has been experimentally confirmed as an effective approach to improve the current-carrying capacity of superconductors, it still does not represent the optimal strategy for enhancing the superconductivity of B(P)SCCO materials.

Metamaterials [34, 35] are artificially engineered microstructures or composite materials that exhibit extraordinary physical properties unattainable in natural materials, characterized by their structure-determined functionalities. Researchers proposed the concept of metamaterial superconductors [36-38], aiming to develop superconducting materials with elevated critical temperatures *Tc* through precisely engineered superconducting architectures. In recent years, the Max Planck Institute proposed light-induced superconductivity [39-43]. Their research team discovered that under mid-infrared laser irradiation, both YBCO, and $K_3C_{60}$ thin films exhibit transient superconducting behavior. Inspired by the concept of metamaterials, our team proposes a novel modification strategy that incorporates electroluminescent heterophases into superconducting materials. By introducing the electroluminescent heterophase as "meta-atoms" into the superconducting particle "matrix", we construct a composite structure. The core mechanism relies on coupling photons generated by electroluminescence from the heterophase with Cooper pairs, thereby reinforcing the pairing strength and enhancing the superconductivity. We define this innovative system as Smart Metamaterial Superconductors (SMSCs).

Previous studies demonstrated that although introducing $Y_2O_3:Eu^{3+}$/Ag luminescent heterophase enhances superconductivity in $MgB_2$ [44-49] ($\Delta Tc$ = +0.4 K, $\Delta Jc$ = +20%) and B(P)SCCO [50-52]($\Delta Tc$ = +1 K, $\Delta Jc$ = +24%), its performance is severely limited by high electric field dependence, low luminescent intensity, and rapid decay characteristics, compromising the enhancement efficiency of superconductor. In contrast, GaN p-n junction particles exhibit superior advantages including high luminescent quantum efficiency and low turn-on voltage, achieving stronger superconducting enhancement in $MgB_2$[53, 54] ($\Delta Tc$ = +0.8 K, $\Delta Jc$ = +35%) and B(P)SCCO[55] ($\Delta Tc$ = +2 K, $\Delta Jc$ = +35%). However, their micron-scale dimensions (average ~2 μm) induce significant impurity effects that substantially offset the potential luminescence gain. Crucially, existing studies only explored the influence of heterophase content on superconducting properties, without systematically investigating the impact of luminescent intensity variation in particles of identical composition on superconducting properties.

Based on the aforementioned work, our team selected GaP quantum dots as the luminescent heterophase for incorporation into $MgB_2$ superconductors[56] . These GaP quantum dots exhibit greater stability than $Y_2O_3:Eu^{3+}$/Ag and smaller dimensions than GaN p-n junction particles. It was observed that both the critical temperature ($Tc$) and critical current density ($Jc$) increased with enhanced luminescent intensity from the GaP quantum dots. In this study, we synthesized $GaP:Zn^{2+}$ / GaP - GaInP - $GaP:Te^{2-}$/GaP core-shell quantum dots [57]with varying electroluminescent intensities via the hot injection method. These quantum dots were introduced as luminescent heterophases into B(P)SCCO superconductors, resulting in a series of modified B(P)SCCO samples. We systematically investigate the influence of both the luminescent intensity and addition content of GaP quantum dots on the superconductivity of B(P)SCCO superconductors

## 2. Experiment

### 2.1 Synthesis of GaP:$Zn^{2+}$ / GaP - GaInP - GaP:$Te^{2-}$/GaP Core-Shell Quantum Dots

In this study, we synthesized GaP:$Zn^{2+}$/GaP-GaInP-GaP:$Te^{2-}$/GaP core-shell quantum dots (hereinafter referred to as GaP QDs) with different distinct electroluminescent intensities via a hot-injection approach through precisely modulating the reaction durations of individual layers (GaP:$Zn^{2+}$/GaP, GaInP, and GaP:$Te^{2-}$/GaP). Figure 1(b) displays the Transmission Electron Microscopy (TEM) image of GaP quantum dots. The inset shows the corresponding size distribution histogram. The sample exhibits a typical tetrahedral nanoparticle morphology of GaP. The nanoparticles demonstrate uniform size distribution, centered at ~3.5 nm (range: 2.5 – 4.5 nm) and following a normal distribution. The slight blurring observed in the imaging primarily originates from the oleylamine ligand layer coating the quantum dot surfaces. The samples underwent X-ray diffraction (XRD) characterization, as shown in Figure 1(c). The vertical black lines denote the standard XRD pattern of zinc-blende structured bulk GaP. As evidenced in the figure, the diffraction peaks of our samples exhibit precise alignment with the reference pattern, demonstrating three characteristic peaks corresponding to the (111), (220), and (311) crystallographic planes of the zinc-blende GaP structure. Notably, no discernible impurity peaks were observed. This demonstrates that the synthesized samples consist of zinc-blende-structured GaP quantum dots, with $Zn^{2+}$ and $Te^{2-}$ ions successfully doped into the GaP crystal lattice without detectable elemental segregation or secondary phase formation. Moreover, the absence of XRD peak shifts demonstrates that low-concentration $Zn^{2+}$/$Te^{2-}$ doping induces no detectable lattice strain. Figure 1(a) displays the electroluminescence (EL) spectrum of the sample under a 7 V bias voltage. The spectrum reveals red emission centered at ~600 nm with a full width at half maximum (FWHM) of ~200 nm under electric field. The 1:1:1 GaP quantum dots (layer-wise reaction time ratio) exhibit a higher emission intensity of ~2950 a.u., while the 1:1.4:1.4 GaP quantum dots show a reduced intensity of ~2100 a.u. Furthermore, we conducted X-ray photoelectron spectroscopy (XPS) analysis to investigate the elemental composition and valence states of the multilayer core-shell GaP nanoparticles. As shown in Figure 1(d)-(f), distinct emission peaks in the Zn 2p, Te 3d, and In 3d characteristic regions confirm the successful doping of $Zn^{2+}$, $Te^{2-}$, and $In^{3+}$ ions into the GaP lattice via the hot-injection method [57].

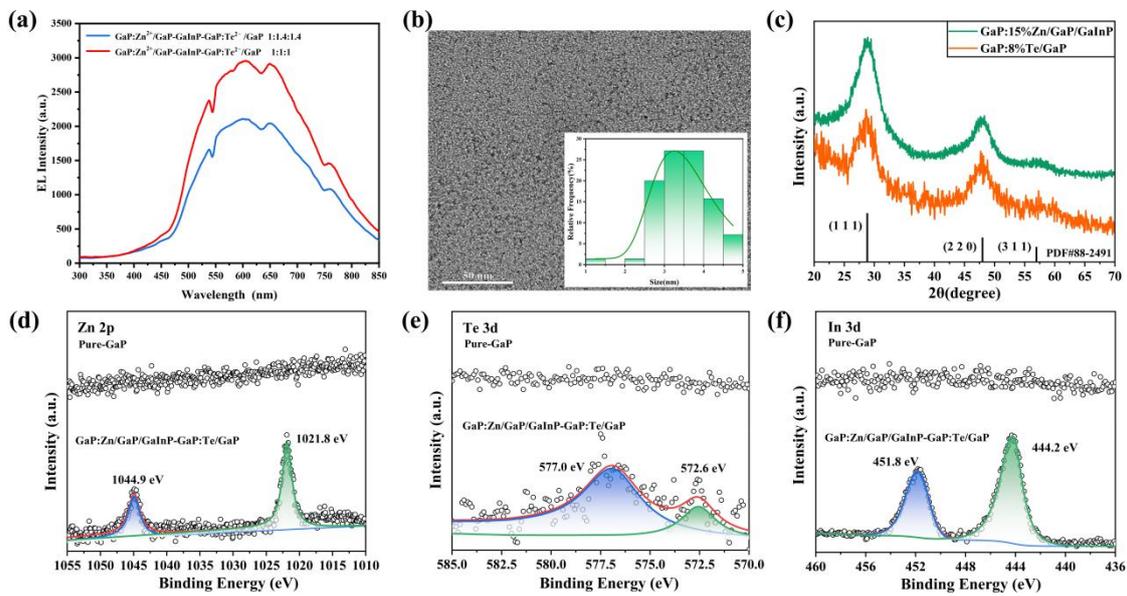

**Figure 1:** (a) EL spectra of GaP quantum dot electroluminescent particles, (b) TEM images of GaP quantum dot electroluminescent particles, (c) XRD patterns of GaP quantum dot electroluminescent particles, and (d), (e), (f) XPS analysis of GaP quantum dot electroluminescent particles.

### 2.2 Preparation of B(P)SCCO Superconducting Samples

In this study, B(P)SCCO superconducting samples were synthesized via solid-state sintering. Stoichiometric quantities of high-purity precursors ($Bi_2O_3$, PbO, $SrCO_3$, $CaCO_3$, CuO; all Alfa Aesar, 99.99%) were weighed according to the molar ratio of 0.9:0.4:2:2:3. The mixed powder and Φ 5 mm agate balls were loaded into a 50 mL milling jar with anhydrous ethanol as a dispersant. Milling was performed in a planetary ball mill (500 rpm, alternating rotation) for 20 h. The as-milled powder was vacuum-dried at 70°C for 60 h, then sieved through a 500-mesh sieve to obtain a uniform particle size of 25 ± 3 μm. The raw material powder was calcined in a tubular furnace in air at 840 °C for 50 hours (heating rate: 4 °C/min; cooling rate: 2 °C/min, with controlled cooling to 800 °C, followed by furnace cooling). After cooling, the powder was collected and subjected to thorough grinding using an agate mortar. This thermal-mechanical treatment cycle was repeated twice, ultimately yielding black pre-calcined B(P)SCCO powder.

A measured amount of B(P)SCCO pre-calcined powder was loaded into a mould and compacted under 12 MPa pressure for 10 minutes to form pellets (Φ 12 × 1 mm). These pellets were then transferred to an alumina boat and sintered in a tubular furnace in air at 840 °C for 30 hours (heating rate: 4 °C/min; cooling rate: 2 °C/min with controlled cooling to 800 °C, followed by furnace cooling). After cooling to room temperature, the pellets were removed and ground thoroughly into B(P)SCCO powder.

**Table 1:** Electroluminescence intensity and addition content of the luminescent hetero-phase of the samples S0 to S4.

| Sample | hetero-phase | Electroluminescence intensity a.u. | Addition content wt.% | Sintering process |
|---|---|---|---|---|
| S0 | / | / | 0 | 840 ℃ 120 h |
| S1 | GaP | / | 0.2 | 840 ℃ 120 h |
| S2 | GaP:$Zn^{2+}$ / GaP - GaInP - GaP:$Te^{2-}$/GaP 1:1.4:1.4 | 2100 | 0.2 | 840 ℃ 120 h |
| S3 | GaP:$Zn^{2+}$ / GaP - GaInP - GaP:$Te^{2-}$/GaP 1:1:1 | 2950 | 0.15 | 840 ℃ 120 h |
| S4 | GaP:$Zn^{2+}$ / GaP - GaInP - GaP:$Te^{2-}$/GaP 1:1:1 | 2950 | 0.2 | 840 ℃ 120 h |

GaP quantum dots with varying electroluminescent intensities were introduced into B(P)SCCO powder at designated addition ratios, mixed by grinding and compacted under 12 MPa pressure for 10 minutes to form pellets (Φ 12 × 1 mm). The pellets were sintered at 840 °C for 120 hours in air (heating rate: 4 °C/min; cooling rate: 2 °C/min with controlled cooling to 800°C, followed by furnace cooling). In this study, the pure

B(P)SCCO sample was designated as S0. Sample containing pure GaP quantum dots without electroluminescence (content: 0.2 wt.%) was designated as S1. Sample with GaP:$Zn^{2+}$/ GaP - GaInP - GaP:$Te^{2-}$/GaP quantum dots added (content: 0.2 wt.%) with a electroluminescence intensity of 2100 a.u. was designated as S2. Samples with GaP:$Zn^{2+}$/ GaP - GaInP - GaP:$Te^{2-}$/GaP quantum dots added (content: 0.15 wt.%, 0.2 wt.%) with a electroluminescence intensity of 2950 a.u. were designated as S3 and S4 respectively, as detailed in Table 1.

### 2.3 Property Characterization of B(P)SCCO Superconducting Samples

The surface morphology of the samples was examined using an FEI Verios G4 ultra-high-resolution field emission scanning electron microscope (SEM). To investigate phase formation, X-ray diffraction (XRD) analysis was performed on a Bruker D8 Advance diffractometer with Cu-Kα radiation (λ = 1.5406 Å) over a 2θ range of 10°–70°. The resistance-temperature (R-T) curve of the sample at low temperatures was measured using the four-probe method to determine the zero-resistance transition temperature $T_{c,0}$ and the onset transition temperature $T_{c,\mathrm{on}}$ of the sample. The entire testing process was conducted under vacuum conditions. The test system was equipped with a closed-cycle cryogenic thermostat manufactured by Advanced Research Systems to provide a cryogenic environment, with the lowest temperature of 10 K. The temperature during the test process was regulated by the Lake Shore 335 cryogenic controller. the test current is provided by the Keithley 6221 source meter, and the test voltage data is collected by the Keithley 2182A digital nanovoltmeter. The critical current density $J_C$ of the samples was determined by measuring the current-voltage (I-V) characteristics under zero magnetic field across various temperatures, allowing for the derivation of the relationship curve between $J_c$ and temperature ($J_c$-T). The DC magnetization measurement is performed in a vacuum low-temperature thermostat. Under zero-field cool-down (ZFC) conditions, a 100 Oe magnetic field is applied, and the Lake Shore 425 Hall magnetometer collects the data of the test magnetic field intensity.

### 3. Results and Discussion

Figure 2(a-e) display the SEM images of samples S0–S4. All specimens exhibit typical quasi-rectangular layered grain structures of the (Bi,Pb)-2223 phase, showing uniform size distribution. Notably, the introduced GaP quantum dot samples (S1-S4) retain intact grain architectures with uniform size distribution comparable to the pure sample S0. This observation demonstrates that the introduction of GaP quantum dot does not compromise the growth of the (Bi,Pb)-2223 high-temperature phase.

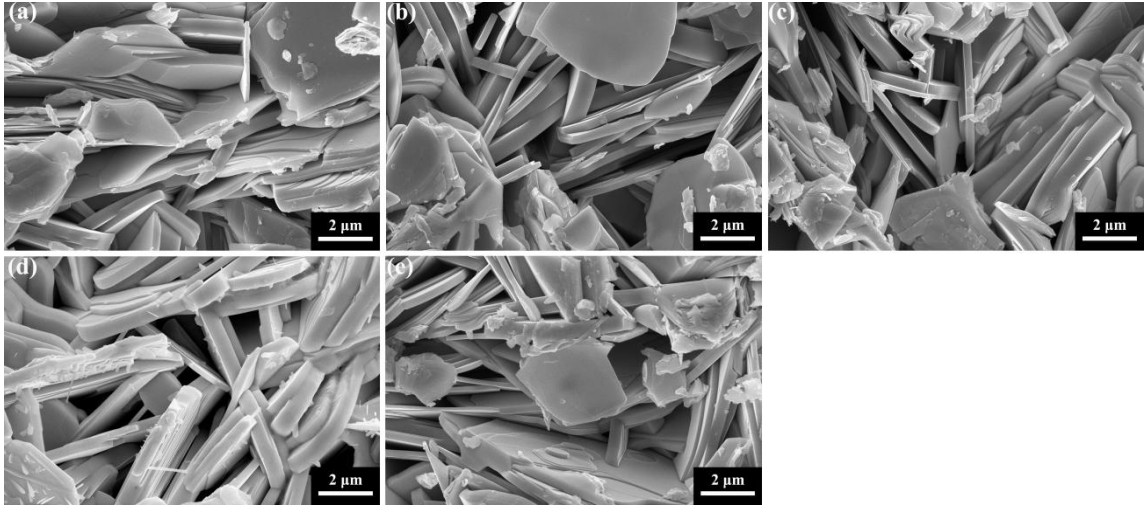

**Figure 2:** SEM images of samples (a) S0, (b) S1, (C) S2, (d) S3, (e) S4.

Phase composition of the samples was characterized by X-ray diffraction (XRD), with corresponding patterns presented in Figure 3. As demonstrated in the figure, all major diffraction peaks correspond to the (Bi,Pb)-2223 high-temperature superconducting phase, while minor peaks originate from the (Bi,Pb)-2212 low-temperature phase. Notably, no oxide peaks associated with Ga, P, In, Te, or Zn were detected, confirming the successful addition of the GaP quantum dot into the B(P)SCCO matrix without decomposition or oxidation during high-temperature sintering. Based on the relative diffraction peak intensities of each phase, we calculated the volume fractions of (Bi,Pb)-2223 and (Bi,Pb)-2212 phase using Eqs. (3-1) and (3-2). The quantified phase fractions are compiled in Table 2, showing no significant variations in the (Bi,Pb)-2223 phase volume fraction across different samples, indicating that the introduction of the GaP quantum dot does not impair (Bi,Pb)-2223 synthesis.

$$(Bi, Pb) - 2212\% = \frac{\sum I(2212)}{\sum I(2223) + \sum I(2212) + \sum I(others)} \times 100\% \quad 3\text{-}1$$

$$(Bi, Pb) - 2223\% = \frac{\sum I(2223)}{\sum I(2223) + \sum I(2212) + \sum I(others)} \times 100\% \quad 3\text{-}2$$

**Table 2:** Lattice parameters and phase volume fractions of the samples S0 to S4.

| Sample | Lattice parameters | | | phase volume fractions | |
|---|---|---|---|---|---|
| | a(Å) | b(Å) | c(Å) | Bi-2223(%) | Bi-2212(%) |
| S0 | 5.407 | 5.416 | 37.118 | 94.49 | 5.51 |
| S1 | 5.407 | 5.416 | 37.118 | 93.35 | 6.65 |
| S2 | 5.407 | 5.416 | 37.118 | 92.49 | 8.51 |
| S3 | 5.407 | 5.416 | 37.118 | 95.39 | 4.61 |
| S4 | 5.407 | 5.416 | 37.118 | 94.16 | 5.84 |

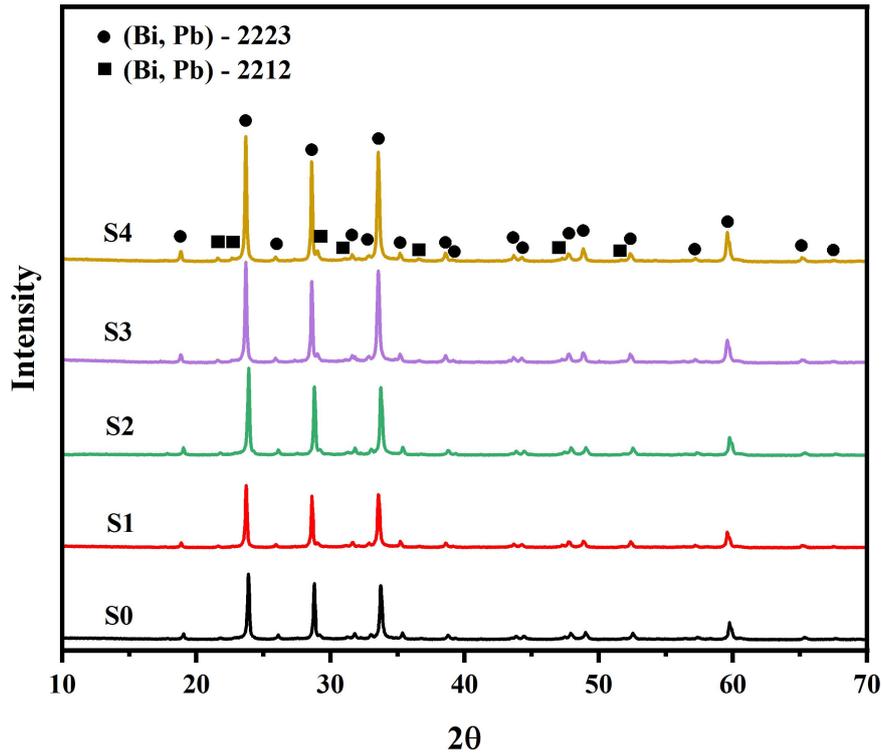

**Figure 3:** XRD patterns of the samples S0 to S4

The resistivity-temperature ($R$–$T$) curves of the samples were measured via standard four-point probe technique in the temperature range of 90–300 K, as presented in Figure 4(a). All samples displayed characteristic superconducting transition behavior: as the temperature gradually decreases near the critical temperature ($T_c$), the resistivity shows a slow decline initially. when approaching the critical temperature $T_c$ from above, the resistivity first manifested a metallic downturn followed by a sharp transition to zero-resistance state.

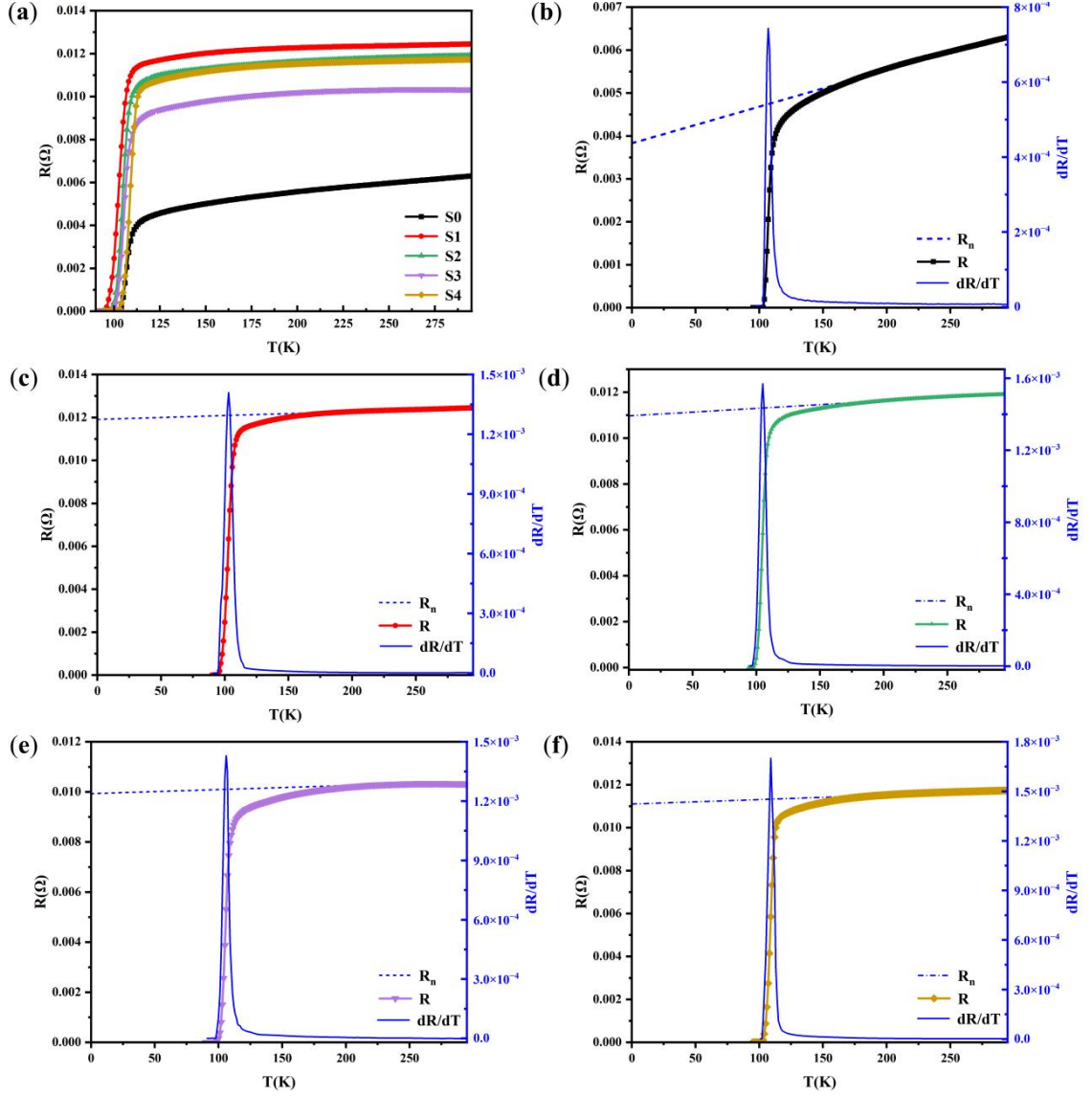

**Figure 4:** (a) The curve of samples resistance-temperature variation within the temperature range of 90-300 K, (b-f) The derivatives (*dR/dT*) of resistance R with respect to temperature *T* at various temperatures (blue solid line), along with the fitted normal-state resistance $R_n$ (blue dashed line) for the samples S0 to S4.

Figure 4(b-f) present the temperature-dependent resistivity *R(T)* and its derivative *dR/dT* (denoted by blue solid curves) for samples S0 – S4. In accordance with Matthiessen's Rule expressed as $R(T) = \alpha T + R_0$, linear regression analysis was performed on the *R(T)* data in the 190–295 K temperature range, enabling determination of three

key parameters: the residual resistance $R_0$, temperature coefficient $α$, and room-temperature resistivity $R_{295K}$. These parameters are systematically summarized in Table 3, with corresponding fitting curves plotted as blue dashed lines in Figure 4(b-f). The temperature-dependent resistivity exhibits metallic behavior during the cooling cycle from room temperature down to $T^*$ (defined via $(R_T−R_0)/αT ≠ 1$), as evidenced by the linear temperature dependence of resistance (Figure 4). When the temperature decreases below $T^*$, the sample enters the pseudogap phase. The resistance gradually deviates from its linear temperature dependence, exhibiting a upward curvature, while $dR/dT$ begins to increase. This phenomenon presumably originates from enhanced electronic correlations accompanying the pseudogap opening, which modify carrier scattering through momentum-dependent self-energy effects. Concurrently, preformed Cooper pairs emerge but lack long-range phase coherence. As the temperature further decreases below the onset transition temperature $T_{c,on}$, the resistance undergoes a steep exponential drop alongside an abrupt rise in $dR/dT$. The sample transitions into the superconducting state, marking the onset of Cooper pair formation. When the temperature decreases below the zero-resistance transition temperature $T_{c,0}$, the sample fully enters the superconducting state, and its resistance plummets to zero. We define the critical transition temperature $T_c$ as the temperature at which the dR/dT value reaches its maximum. The onset transition temperature ($T_{c,on}$), zero-resistance transition temperature ($T_{c,0}$), and critical transition temperature ($T_c$) for all samples are summarized in Table 3.

The transition temperature ranges for each sample are: S0 103–114K, S1 95–112K, S2 97 K–113 K, S3 98K–114K, S4 102K– 114K. The pure sample S0 exhibits a critical temperature $T_c$ of 107 K. For the GaP quantum dot-added samples S1–S4, the critical temperatures $T_c$ are measured as 104 K, 105 K, 106 K, and 109 K, respectively. As evidenced by samples S1, S2, and S4 with identical heterophase addition levels (0.2 wt.%), the non-luminescent GaP quantum dot added sample S1 exhibits a 3 K reduction in critical transition temperature $T_C$ compared to the pure sample. Notably, $T_C$ progressively increases with enhanced electroluminescence intensity of the introduced GaP quantum dot. In samples S3 and S4 with heterophase additions of identical electroluminescence intensity, Sample S4 with a 0.2 wt.% heterophase addition demonstrates a higher critical transition temperature ($T_c$) compared to sample S3 containing a 0.15 wt.% of GaP quantum dot, while exhibiting a 2 K enhancement relative to the pure sample. These findings demonstrate that the electroluminescent heterophase exerts an enhancement effect on the superconductivity of B(P)SCCO. This enhancement correlates positively with both the electroluminescence intensity and the addition content of the heterophase, where stronger luminescence intensity yields more pronounced superconducting property improvements. As evidenced in Table 3, the heterophase-added samples (S1-S4) exhibit room-temperature resistivity ($R_{295K}$) and fitted residual resistivity ($R_0$) values that are 1.5 to 2 times higher than those of the pure sample S0. Concurrently, these modified samples S1–S4 demonstrate lower temperature coefficient α values compared to the pure sample S0, indicating more gradual resistivity variations within the 160–300 K range. This observation confirms that heterophase addition in B(P)SCCO superconductors introduces impurity effects that perturb charge transport properties. These impurity effects compete with the electroluminescence-induced enhancement effects, collectively affecting the superconducting behavior. Furthermore, in samples containing non-luminescent or low-electroluminescence heterophase additions, the impurity effect dominates over the enhancement effect, resulting in lower critical transition temperatures ($T_c$) compared to the pure sample.

**Table 3:** Critical transition temperature $T_C$, room-temperature resistance $R_{295K}$ and fitted residual resistance $R_0$ of the samples S0 to S4.

| Sample | $T_{c,0}$(K) | $T_{c,on}$(K) | $T_c$(K) | $T^*$(K) | $R_{295K}$(Ω) | α | $R_0$(Ω) |
|---|---|---|---|---|---|---|---|
| S0 | 103 | 114 | 107 | 160.2 | 0.00630 | 8.52e−6 | 0.00383 |
| S1 | 95 | 112 | 104 | 163.8 | 0.01190 | 1.80e−6 | 0.01216 |
| S2 | 97 | 113 | 105 | 161.4 | 0.01097 | 3.25e−6 | 0.01097 |
| S3 | 98 | 114 | 106 | 186.9 | 0.01032 | 1.62e−6 | 0.00993 |
| S4 | 102 | 114 | 109 | 169.1 | 0.01173 | 2.08e−6 | 0.01113 |

Figure 5 presents the temperature-dependent critical current density ($J_c$-$T$) curves of the samples. The data demonstrate that the critical current density follows the characteristic temperature evolution of B(P)SCCO superconductors: $J_c$ decreases monotonically with increasing temperature until reaching the threshold where superconductivity is fully suppressed, at which point $J_c$ approaches zero. Samples S1–S4 incorporating the GaP quantum dot exhibit accelerated degradation of critical current density ($J_c$) compared to the pure sample. This is because the impurity effect caused by the introduction of the heterostructure leads to a lower zero-resistance transition temperature $T_{c,0}$ of the sample compared to the pure sample, enabling the sample to enter a transition state from the superconducting state to the pseudogap phase. As evident from the figure, under isothermal conditions, samples S1, S2, and S4—each containing identical GaP quantum dot content—exhibit a monotonic increase in critical current density ($J_c$) concurrent with the amplification of electroluminescence intensity. Compared with sample S3, sample S4 with higher content of GaP hetero-phase has a higher critical current density $J_c$, and it shows a 20% higher $J_c$ than pure sample S0. This demonstrates that the electroluminescence effect of GaP quantum dot can effectively enhance the critical current density ($J_c$) of B(P)SCCO superconductors, with this enhancement being positively correlated with both the electroluminescence intensity and the added content of the luminescent GaP quantum dot.

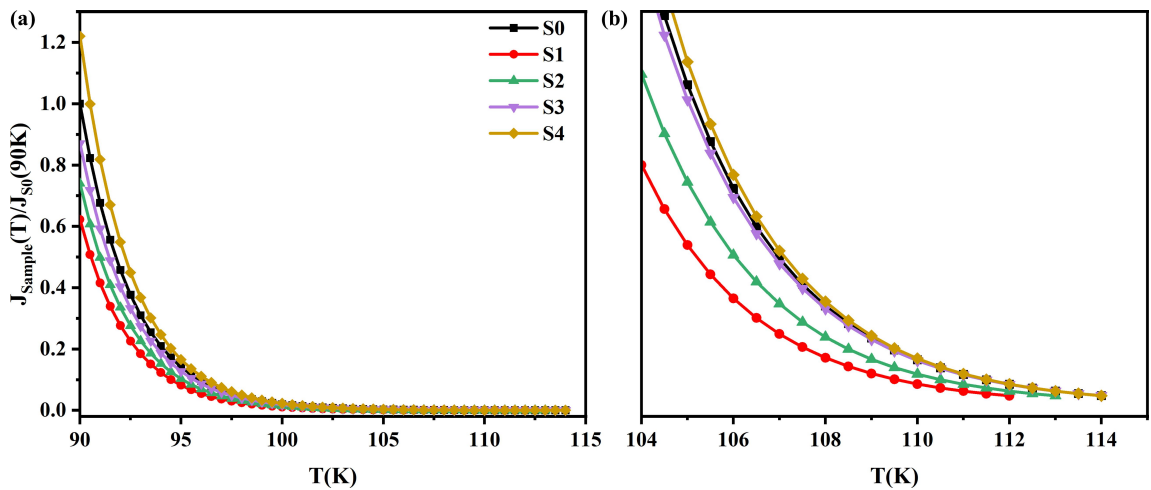

**Figure 5:** (a) Critical current density $J_c$ of the samples S0 to S4 within the temperature range of 90-115 K, (b) Partially enlarged view of $J_C$-$T$ curve.

Figure 6 shows the temperature-dependent magnetic susceptibility of the sample under a 100 Oe magnetic field in zero-field-cooled (ZFC) mode. All four curves exhibit a superconducting diamagnetic transition, where the diamagnetism reaches its maximum at low temperatures and progressively diminish with increasing temperature until disappearing completely. This trend aligns with the characteristic evolution of B(P)SCCO superconductors, confirming integrity of diamagnetic properties in all four specimens. The diamagnetic responses of samples S0, S1, S2, S3, and S4 vanish at 107K, 102K, 104.5K, 105K, and 108K, respectively. The experimental results demonstrate that the introduction of GaP luminescent quantum dot effectively enhances the superconducting properties of B(P)SCCO, with this enhancement being positively correlated with both the electroluminescence intensity and the concentration of the addition of the heterophase.

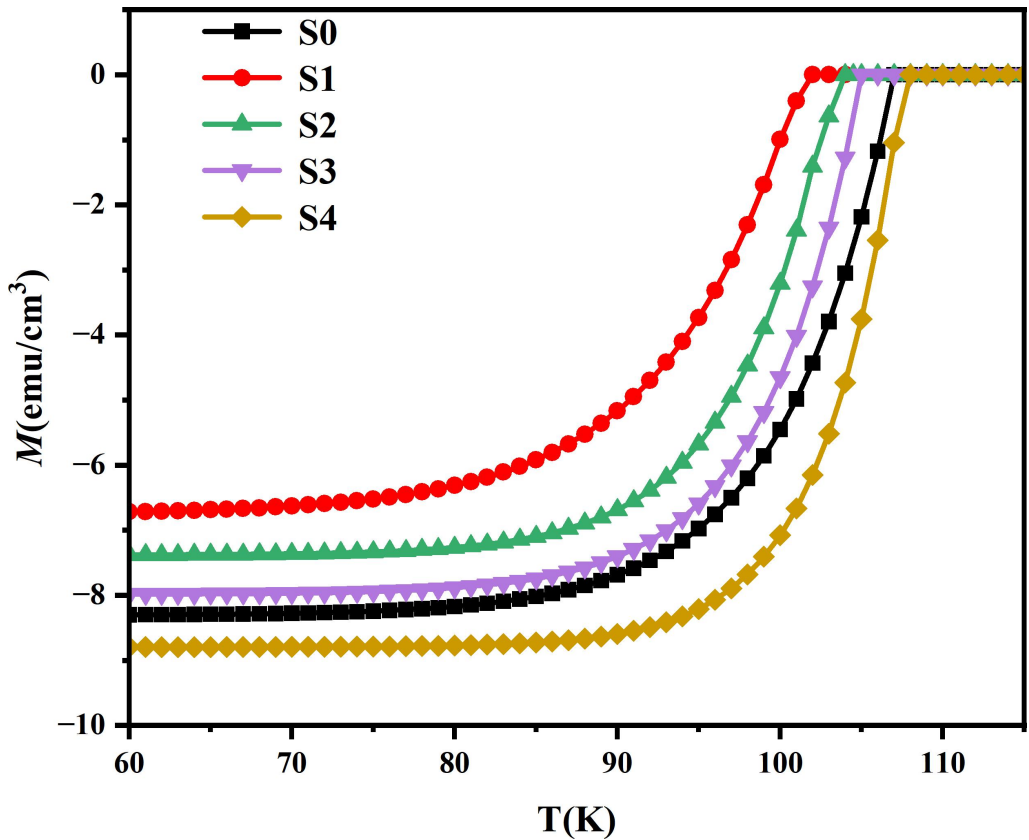

**Figure 6:** The DC magnetic susceptibilities of the samples S0 to S4.

In summary, the experimental results demonstrate that introducing electroluminescent GaP heterophases modulates the superconductivity of B(P)SCCO. Compared to the pristine sample (S0), sample S1 with non-luminescent GaP quantum dots exhibits a 3 K reduction in critical temperature ($T_c$) and a 37% decrease in critical current density ($J_c$), indicating that GaP quantum dots induce impurity effects that degrade superconducting properties. As the luminescent intensity of GaP quantum dots increases, samples S2 and S3 show progressive recovery in $T_c$ and $J_c$ relative to S1. This reveals that the electroluminescence of GaP quantum dots generates an enhancement effect, which scales with luminescent intensity and competes with impurity effects, ultimately leading to gradual improvement in superconductivity. Further increasing the addition content of luminescent GaP quantum dots (sample S4) not only continues to enhance $T_c$ and $J_c$ but also surpasses the pristine sample (S0), achieving a $T_c$ increase of 1 K and a $J_c$ enhancement of 20%. This indicates that the electroluminescence-induced enhancement depends on both luminescent intensity and quantum dot concentration. When electroluminescent intensity and addition content are sufficiently high, the enhancement effect can fully overcome impurity effects, resulting in a net gain in superconducting performance.

Furthermore, owing to the inherent structural stability of GaP quantum dots (resistance to degradation), their electroluminescent intensity exhibits exceptional stability. Experimental verification confirms that the GaP quantum dots prepared in this work show no significant attenuation in luminescent intensity after six months of storage. More importantly, the B(P)SCCO superconductor samples doped with these GaP quantum dots, when stored under identical conditions for six months, demonstrate critical temperature ($T_c$) and critical current density ($J_c$) values comparable to initial measurements, with no observable deterioration. These results demonstrate that the incorporation of GaP quantum dots not only ensures stable coexistence within the B(P)SCCO matrix but also delivers long-term stability in enhancing superconducting properties (e.g., $T_c$ and $J_c$). The enhancement effect exhibits negligible decay over extended periods.

Unlike traditional nano-compound particles that exhibit "monofunctional regulation" of B(P)SCCO superconductor performance (solely enhancing critical current density $J_c$), the introduction of luminescent GaP quantum dots demonstrates a unique "dual enhancement" — simultaneously elevating both critical temperature ($T_c$) and $J_c$. Nano-compound particles primarily enhance $J_c$ by introducing flux pinning centers to suppress flux motion. However, their addition typically reduces Tc due to lattice mismatch or interface scattering that impedes carrier transport. Their modification efficacy depends solely on nanoparticle concentration. In contrast, GaP quantum dots, as an heterophases, participate in superconducting state modulation primarily through their electroluminescent properties. Their dual enhancement effect is simultaneously dependent on both the addition content and luminescent intensity of the quantum dots. We propose a physical model wherein photons generated via electric-field excitation of the GaP luminescent inhomogeneous phase interact with superconducting electrons, triggering surface plasmon formation. These evanescent surface waves[53, 58] enable coherent transport of energy-matched superconducting electrons, amplifying electronic interactions within the plasmonic system. The coupling between electroluminescence-induced evanescent waves and superconducting electrons leads to smart superconductivity. Experimental evidence confirms that the resonant enhancement of superconducting electron transport properties exhibits a direct positive correlation with electroluminescence intensity.

## 4. Conclusion

This study systematically investigates the "dual-effect modulation" mechanism of GaP luminescent quantum dots on the critical temperature ($T_c$) and critical current density ($J_c$) of B(P)SCCO superconductors. GaP quantum dots with controllable luminescent intensity were synthesized via the hot injection method and incorporated into B(P)SCCO superconductors. XRD and SEM characterizations confirm that their introduction neither alters the (Bi,Pb)-2223 phase content nor compromises the dense and uniform microstructure of B(P)SCCO. Electrical transport measurements reveal that introducing GaP quantum dots with high luminescent intensity and optimal addition content elevates $T_c$ by 1 K and enhances $J_c$ by 20%. In contrast, samples with low-luminescence or non-luminescent GaP quantum dots exhibit reduced $T_c$ and $J_c$. This demonstrates that the enhancement of superconductivity by GaP quantum dots depends on both their luminescent intensity and addition content, while simultaneously introducing impurity effects that compete with the enhancement mechanism. This discovery provides a novel strategy for optimizing superconductor performance. This study holds significant practical value for superconducting applications: In superconducting cables, a higher critical current density ($J_c$) enables either greater current-carrying capacity at identical cross-sections or reduced cable dimensions while maintaining equivalent current loads. For superconducting magnets, the enhancement in $J_c$ directly translates to stronger magnetic field generation capabilities. Crucially, the incorporation of GaP quantum dots preserves the (Bi,Pb)-2223 phase content and microstructure, ensuring processing stability during material fabrication.

**Acknowledgments:** This research was supported by the National Natural Science Foundation of China for Distinguished Young Scholar under Grant No. 50025207.

**Declaration of competing interest:** We declare that they have no known competing financial interests or personal relationships that could have appeared to influence the work reported in this paper.

**Author Contributions**: Conceptualization, methodology, X.Z.; software, Q.H., D.C.; validation, Q.H., D.C. and Y.Q.; formal analysis, Q.H., D.C. and R.B.; investigation, Q.H., D.C., Y.Q., X.L. and X.L.; resources, X.Z.; data curation, Q.H., D.C. and Y.Q., writing—original draft preparation, Q.H.; writing—review and editing, Q.H. and X.Z.; visualization, Q.H., D.C. and R.B.; supervision, X.Z.; project administration, X.Z.; funding acquisition, X.Z. All authors have read and agreed to the published version of the manuscript.

First Author and Second Author contribute equally to this work.

**Institutional Review Board Statement:** Not applicable.

**Data Availability Statement:** The data presented in this study are available on reasonable request from the corresponding author.